\documentclass{iau307}
\usepackage{graphicx}
\usepackage{natbib}
\usepackage{url}
\usepackage{dtklogos}
\bibpunct{(}{)}{;}{a}{}{,}

\title[Pulsational Waves Probed by Photospheric Emission] {Deep Photospheric
  Emission Lines as Probes for Pulsational Waves}

\author[Th.~Rivinius, M.~Shultz, G.A.~Wade]
{Th.~Rivinius$^{1}$ \and M.~Shultz$^{1,2,3}$ \and  G.\,A.~Wade$^{3,2}$}

\affiliation{$^{1}$ESO - European Organisation for Astronomical Research in the Southern Hemisphere, Chile\\ email: {\tt triviniu@eso.org} \\[\affilskip]
$^{2}$Dept. of Physics, Engineering Physics and Astronomy, Queen's University, Canada\\[\affilskip]
$^{3}$Dept. of Physics, Royal Military College of Canada, Canada

}

\pubyear{2014}
\volume{307} 
\pagerange{}
\jname{New windows on massive stars: asteroseismology, interferometry, and spectropolarimetry}
\editors{G. Meynet, C. Georgy, J.H. Groh \& Ph. Stee, eds.}

\begin{document}

\maketitle

\begin{abstract}
Weak line emission originating in the photosphere is well known from O stars
and widely used for luminosity classification. The physical origin of the line
emission are NLTE effects, most often optical pumping by far-UV lines.
Analogous lines in B stars of lower luminosity are identified in radially
pulsating $\beta$ Cephei stars. Their diagnostic value is shown for radially
pulsating stars, as these lines probe a much larger range of the photosphere
than absorption lines, and can be traced to regions where the pulsation
amplitude is much lower than seen in the absorption lines.
\end{abstract}

\firstsection 
\section{Introduction}
Weak {line emission originating in the photosphere} is well known from
luminous O stars. The physical origin of the line emission is NLTE effects,
most often optical pumping by far-UV lines. In early B hypergiants Fe{\sc iii}
emission lines were identified to be pumped by He{\sc i} lines
\citep{1985A&A...148..412W}. \citet{1997A&A...318..819R} found these lines to
have the least negative radial velocity and the least degree of
variability. This means they are {seated deeply in the photosphere}.
Analogous lines can be found in early B-type $beta$ Cephei stars of lower
luminosity. Their {diagnostic value} is shown for radially pulsating stars, as
these lines {probe a much larger range of the photosphere} than absorption
lines.

\begin{figure}[t]
\begin{center}
\includegraphics[width=\textwidth]{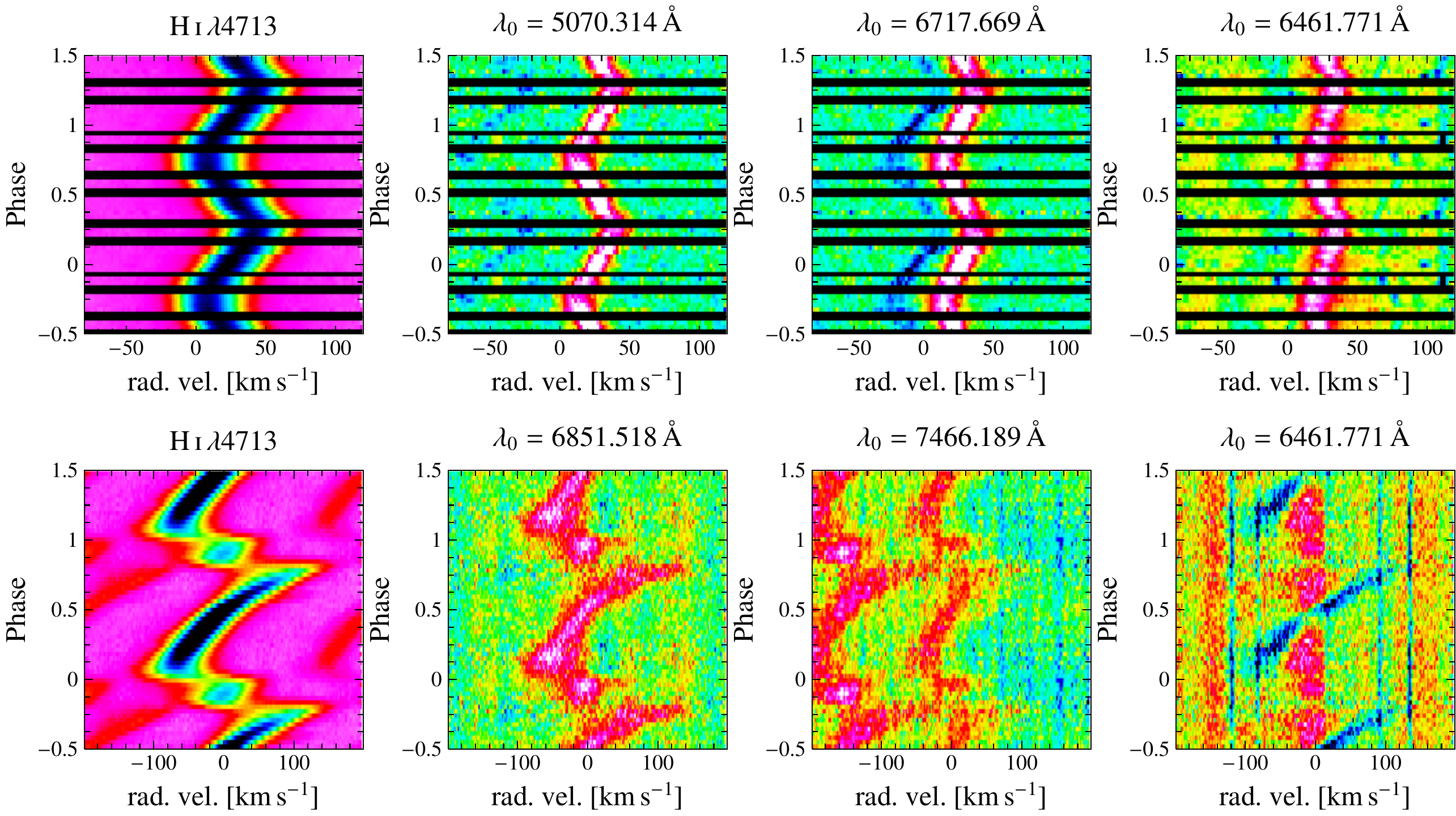}
\caption{Phased $\beta$\,Cephei-variability in $\xi^1$\,CMa (top) and BW\,Vul
  (bottom) in selected lines.  Absorption lines, like the rightmost, all have
  similar amplitudes. The lower amplitude in some emission lines (three
  leftmost panels in each row) is evident.}
\label{Rivinius_bceps_fig1}
\end{center}
\end{figure}

\section{Observations}

Observations of $\xi^1$\,CMa (116 proprietary intensity spectra) and BW\,Vul
(434 archival intensity spectra) were carried out with {ESPaDOnS/CFHT} on
Mauna Kea ($R=68\,000$, 370 to 1000\,nm, $S/N>100$ for all observations). Data
were usually obtained by continuously observing the targets over several
hours.

For $\xi^1$\,CMa the formation of the weak emission lines in the photosphere is
corroborated by their magnetic signature, which is identical in shape, but
{inverted in sign w.r.t.\ the absorption lines. This is the {expected
    behaviour for photospherically formed emission}. See Shultz et al.\ (this
  volume) for a more in-depth discussion of the magnetic properties of
  $\xi^1$\,CMa

For most lines, the pumping or fluorscence mechanisms were not yet identified.
However, for several of the emission lines that could be identified, the upper
levels coincide with the upper levels of strong potassium resonance lines
below about 100\,nm. The lines in which emission is seen were tentatively
identified as
\begin{itemize}
\item {\bf Si{\sc ii}:} $\lambda\lambda$ 6347, 6371. 
\item {\bf Si{\sc iii}:}
$\lambda\lambda$ 6462 (or C{\sc ii}), 7463, 7467, 8103.
\item {\bf Fe{\sc iii}:} $\lambda$ 5879.
\item {\bf C{\sc ii}:} $\lambda\lambda$ 6151, 6462 (or Si{\sc iii})
\item {\bf Unidentified:} $\lambda\lambda$ 5056, 5070, 6587, 6718, 6777, 6804,
6806, 6852, 6928, 7113, 7116, 7388, 7552, 7512, 7852, 8236, 8287, 8293, 8630,
9826, 9855.
\end{itemize}

\section{Discussion}

Fig.~\ref{Rivinius_bceps_fig1} shows that the pulsation amplitude in the
emission lines is either similar or smaller than in the absorption lines.  It
is known for radial pulsators that there is an amplitude gradient in the
photosphere. For high amplitude pulsators, the amplitude increases with height
\citep[see Fig.\ 2 of][for models]{2004A&A...426..687F}.

However, the typical change of amplitude is rather small over the range of the
photospheric absorption lines. Fig.\ 6 of \citet{2013A&A...553A.112N}
indicates changes of up to 10\% between weak and strong absorption lines. In
the case of $\xi^1$\,CMa, the emission lines with the lowest peak-to-peak
amplitudes have an {amplitude of only 50\%} of the amplitude of the absorption
lines, while for BW\,Vul the {amplitude seems even to approach zero} in the
most extreme case (see lower row of {Fig.~\ref{Rivinius_bceps_fig1}}, He{\sc
  i}$\lambda$4713 vs. $\lambda_0=6461.771$\,\AA).

\bibliographystyle{iau307}
\bibliography{bceps}

\end{document}